\pdfoutput=1

\documentclass[aps,10pt,twocolumn,showpacs,superscriptaddress,footnoteinbib]{revtex4}
\usepackage{amsmath}
\usepackage{latexsym}
\usepackage{amssymb}
\usepackage{graphics,epstopdf}

\begin{document}
\title{Construction of optimal teleportation witness operators from entanglement witnesses}
\author{Satyabrata Adhikari}
\thanks{satya@iitj.ac.in}
\affiliation{Indian Institute of Technology Rajasthan, Jodhpur-342011, India}
\author{Nirman Ganguly}
\thanks{nirmanganguly@gmail.com}
\affiliation{Department of Mathematics, Heritage Institute of Technology, Kolkata-700107,  India}
\affiliation{S. N. Bose National Centre for Basic Sciences, Salt Lake,
Kolkata-700098, India}
\author{A. S. Majumdar}
\thanks{archan@bose.res.in}
\affiliation{S. N. Bose National Centre for Basic Sciences, Salt Lake,
Kolkata-700098, India}
\date{\today}
\begin{abstract}
Teleportation witnesses are hermitian operators which can identify useful entanglement for quantum teleportation. Here we provide a systematic method to construct teleportation witnesses from entanglement witnesses corresponding to general qudit systems. The witnesses so constructed are shown to be optimal for qubit 
and qutrit systems, and therefore detect the largest set of states useful for teleportation within a given class. We demonstrate the action of the witness pertaining to different classes of states in qubits and qutrits.  Decomposition of the witness in terms of spin operators facilitiates experimental identification of useful resources for teleportation. 
\end{abstract}


\maketitle

\section{INTRODUCTION}

Entanglement has long been identified as a unique feature of quantum states due
 to the seminal work by Schrodinger \cite{schro} and EPR \cite{epr}. The notion 
of quantum entanglement has been extensively studied \cite{horoent} and has 
paved the way for modern quantum information science \cite{nielsen} enabling 
tasks such as teleportation, superdense coding and cryptography\cite{teleport,dense,crypto}, which are beyond the reach of classical physics.
Since entanglement is the essential ingredient for several quantum information
processing tasks,  its detection is important. Experimental detection of 
entanglement is made possible by entanglement witnesses \cite{horowit,terhalwit}. Based upon the Hahn-Banach separation axiom from functional analysis \cite{holmes}, entanglement witnesses serve to demarcate entangled states from the ones 
which are separable. Entanglement witnesses \cite{sperling1,ganguly1,reviewwit}
 provide a necessary and sufficient entanglement criterion in terms of directly measurable observables \cite{reviewwit,guhne,barbieri,hyllus,bertlmann} 
facilitating experimental detection of entanglement. 

Entanglement witnesses are
not universal, and hence the  question as to  how to maximally detect entangled
 states, i.e.,  increase the number of states detected by the witness, is 
of significance. The possibility of optimization of entanglement witnesses
\cite{optwit} has lead to the  construction of optimal witnesses \cite{bertl,chruscinski}. The study of entanglement witnesses has proceeded also in the 
direction of Schmidt number witnesses \cite{terschmidt,sanpera,sperling2} and 
common witnesses \cite{guo,ganguly2}. 
Entanglement witnesses enable experimentally viable procedures to detect the 
presence of entanglement, a notion that has been carried forward to identify 
manifestations of various properties of quantum states, such as macroscopic 
entanglement through 
thermodynamical witnesses\cite{thermowit}, as well as witnesses for quantum 
correlations \cite{discordwit}, teleportation \cite{ganguly3,zhao}, 
cryptography \cite{cryptowit} and mixedness 
\cite{mixedwit}. 

 Although entanglement is a key ingredient for teleportation, yet not all 
entangled states are useful for the purpose of teleportation. The problem
 gets accentuated in higher dimensions where bound entangled states 
\cite{bound1} are also present. The ability of an entangled state to perform 
teleportation is linked to a threshold value of the fully entangled 
fraction \cite{fef1} which is difficult to estimate except for some known states \cite{fef3}. Based upon the linkage of the threshold value of the fully
entangled fraction with teleportation fidelity, and utilising again the 
separation axioms, the existence of hermitian operators acting as teleportation
witness was demonstrated recently \cite{ganguly3}.  A teleportation witness $W_{T}$ is a hermitian operator with at least one negative eigenvalue and (i) $Tr(W_{T}\varpi)\geq 0$, for all states $\varpi$ not useful for teleportation and (ii) $Tr(W_{T}\vartheta)<0$ for atleast, one entangled state $\vartheta$ which is useful for teleportation. In a following work \cite{zhao} a teleportation witness
with interesting universal properties was proposed, which though depends upon 
the choice of a unitary operator that may be difficult to find in practice, 
especially in higher dimensions.

The difficulty in identifying useful resources for teleportation necessitates 
the construction of suitable teleportation witnesses that would be possible to 
implement experimentally in order to ascertain whether a given unknown state 
would be useful as a teleportation channel. Moreover, analogous to the theory 
of entanglement 
witnesses, maximal detection of states capable for teleportation is a question 
of significance. The motivation of this work is to  address both the above 
issues. In the present paper we propose an efficient method to construct 
teleportation witnesses for general qudit systems starting from entanglement 
witnesses. We next 
demonstrate the optimality of such a witness for the case of qubit and qutrit 
systems,   exemplifying its action for different classes of states. We 
further decompose the witness in terms of spin operators, thereby taking a
step towards the viability for its experimental realization.

\section{OPTIMAL TELEPORTATION WITNESS}

Amongst two witnesses $W_{1}$ and $W_{2}$, $W_{1}$ is said to be finer than $W_{2}$, if $DW_{2}\subseteq DW_{1}$, where $DW_{i}=\lbrace \chi: Tr(W_{i}\chi)<0\rbrace, i= 1,2$, i.e., the set of entangled states detected by $W_{i}$. A witness is 
said to be optimal if there exists no other witness finer than it \cite{optwit}. Further, if the set of product vectors $\vert e,f\rangle$,  $P_{W}=\lbrace \vert e,f\rangle: Tr(W\vert e,f\rangle \langle e,f\vert)=0 \rbrace$, spans the 
relevant product Hilbert space, then the witness $W$ is optimal \cite{optwit}. 
Recently, it was shown \cite{lewenstein} that if a witness operating on $H_{m}\otimes H_{m}$ can be expressed in the form $W=Q^{T_{A}}$, where $Q$ is the projector on a pure entangled state, then the witness $W$ is optimal.

On the other hand, as stated earlier, the ability of a quantum state in performing teleportation is determined by a threshold value of the fully entangled fraction, given by  $F(\rho)=max_{U}Tr[(U^\dagger \otimes I)\rho (U \otimes I)\vert \Phi \rangle \langle \Phi \vert ]$ \cite{fef1}, where $ \vert \Phi \rangle=\frac{1}{\sqrt{d}}\sum _{k=0}^{d-1} \vert kk \rangle$ and $U$ is an unitary operator. Precisely, in $d \otimes d$ systems if $F(\rho)$ exceeds $\frac{1}{d}$, then the state is considered useful for the protocol \cite{fef1}. 

\subsection{Optimal teleportation witness for qubits}

Consider the entanglement witness, $W^{2}=\rho_{\phi^{+}}^{T_{A}}$, where $\vert \phi^{+} \rangle=\frac{1}{\sqrt{2}}(\vert 00 \rangle + \vert 11 \rangle) $, acting on two qubit systems. Since, $\rho_{\phi^{+}}= \frac{1}{4}(I \otimes I + \sigma_{x} \otimes \sigma_{x} - \sigma_{y} \otimes \sigma_{y}+\sigma_{z} \otimes \sigma_{z})$, one thus obtains,
$ W^{2}=\frac{1}{4}(I \otimes I + \sigma_{x} \otimes \sigma_{x} + \sigma_{y} \otimes \sigma_{y}+\sigma_{z} \otimes \sigma_{z})$, which implies,
\begin{eqnarray}
&&Tr((W^{2}-\frac{1}{4}\sigma_{y} \otimes \sigma_{y})\rho)\nonumber\\
&&=\frac{1}{4}Tr((I \otimes I + \sigma_{x} \otimes \sigma_{x} +\sigma_{z} \otimes \sigma_{z})\rho)\label{witpauli}
\end{eqnarray}
for any arbitrary density matrix $\rho$. Hence,
\begin{eqnarray}
F(\rho)\geq Tr(\rho \vert \phi^{+}\rangle \langle \phi^{+} \vert)
\label{witpauli2}
\end{eqnarray}
The r.h.s of the above equation is given by 
$\frac{1}{4}Tr((I \otimes I + \sigma_{x} \otimes \sigma_{x} +\sigma_{z} \otimes \sigma_{z}-\sigma_{y} \otimes \sigma_{y})\rho)$, which 
using Eq.(\ref{witpauli}),  becomes
$Tr((W^{2}-\frac{1}{2}\sigma_{y}\otimes\sigma_{y})\rho)$.
 This in turn implies using Eq.(\ref{witpauli2}) that
\begin{equation}
Tr((\frac{1}{2}\sigma_{y}\otimes\sigma_{y}+\frac{1}{2}I-W^{2})\rho)\geq \frac{1}{2}-F(\rho)
\end{equation}
If $\rho$ is not useful for teleportation, i.e., $F(\rho)\leq \frac{1}{2}$, 
then $Tr((\frac{1}{2}\sigma_{y}\otimes\sigma_{y}+\frac{1}{2}I-W^{2})\rho)\geq 0$,
 implying that 
\begin{equation}
W_{2 \otimes 2}= \frac{1}{2}\sigma_{y}\otimes\sigma_{y}+\frac{1}{2}I-W^{2} \label{telqubit}
\end{equation}
is a teleportation witness acting on two qubits.

Next, with some straightforward algebraic manipulation it is observed that the 
witness can be expressed as
\begin{equation}
W_{2 \otimes 2}=(\vert \psi^{-}\rangle \langle \psi^{-}\vert)^{T_{A}}
\end{equation}
where, $\vert \psi^{-} \rangle = \frac{1}{\sqrt{2}}(\vert 01 \rangle - \vert 10 \rangle)$. Further the product vectors
$(\vert 0 \rangle + i\vert 1\rangle) \otimes (\vert 0 \rangle - i\vert 1\rangle), (\vert 0 \rangle + \vert 1\rangle)^{\otimes 2}, \vert 00 \rangle , \vert 11 \rangle$ span $C^{2} \otimes C^{2}$ and belong to $P_{W_{2 \otimes 2}}$.
This establishes the optimality of the teleportation witness \cite{optwit,lewenstein}.

\subsection{Optimal teleportation witness for qutrits}

The generalized Gell-Mann matrices are higher
dimensional extensions of the Pauli matrices (for qubits) and are
hermitian and traceless. They form an orthogonal set and
basis. In particular, they can be categorized for qutrits as the following types of traceless matrices \cite{bertlmann}:
\begin{align}
\lambda^{1}= \left(%
\begin{array}{ccc}
  0 & 1 & 0 \\
  1 & 0 & 0 \\
  0 & 0 & 0 \\
\end{array}%
\right), & \lambda^{2}= \left(%
\begin{array}{ccc}
  0 & \text{-i} & 0 \\
  \text{i} & 0 & 0 \\
  0 & 0 & 0 \\
\end{array}%
\right),  & \lambda^{3}= \left(%
\begin{array}{ccc}
  1 & 0 & 0 \\
  0 & -1 & 0 \\
  0 & 0 & 0 \\
\end{array}%
\right), \nonumber
\end{align}
\begin{eqnarray}
\lambda^{4}=
\left(%
\begin{array}{ccc}
  0 & 0 & 1 \\
  0 & 0 & 0 \\
  1 & 0 & 0 \\
\end{array}%
\right), \lambda^{5}=
\left(%
\begin{array}{ccc}
  0 & 0 & \text{-i} \\
  0 & 0 & 0 \\
  \text{i} & 0 & 0 \\
\end{array}%
\right),\lambda^{6}=
\left(%
\begin{array}{ccc}
  0 & 0 & 0 \\
  0 & 0 & 1 \\
  0 & 1 & 0 \\
\end{array}%
\right),
\nonumber
\end{eqnarray}
\begin{eqnarray}
\lambda^{7}=
\left(%
\begin{array}{ccc}
  0 & 0 & 0 \\
  0 & 0 & \text{-i} \\
  0 & \text{i} & 0 \\
\end{array}%
\right),\lambda^{8}=
\left(%
\begin{array}{ccc}
  1/\sqrt{3} & 0 & 0 \\
  0 & 1/\sqrt{3} & 0 \\
  0 & 0 & -2/\sqrt{3} \\
\end{array}%
\right)\nonumber
\end{eqnarray}
Now, consider the following entanglement witness in qutrits,
\begin{equation}
W^{3}= (\vert \delta \rangle \langle \delta \vert)^{T_{A}}
\end{equation}
where, $\delta=\frac{1}{\sqrt{3}}(\vert 00 \rangle + \vert 11 \rangle + \vert 22 \rangle)$,  yielding,
\begin{eqnarray}
W^{3}&&=\frac{1}{9}(I \otimes I + \frac{3}{2}\Delta) 
\end{eqnarray}
with $\Delta= \sum_{i=1}^{8}\lambda^{i} \otimes \lambda^{i}$.
Therefore, for any arbitrary density matrix $\sigma \in B(H_{3} \otimes H_{3})$ ,taking $\Delta_{1}=\lambda^{2}\otimes \lambda^{2}+\lambda^{5}\otimes \lambda^{5}+\lambda^{7}\otimes \lambda^{7}$ and $\Delta_{2}=\lambda^{1}\otimes \lambda^{1} +\lambda^{3}\otimes  \lambda^{3}+\lambda^{4}\otimes \lambda^{4} +\lambda^{6}\otimes \lambda^{6}+\lambda^{8}\otimes \lambda^{8}$, one gets
\begin{eqnarray}
Tr[(W^{3}-\frac{1}{6}\Delta_{1})\sigma]
= \frac{1}{9}Tr[(I \otimes I + \frac{3}{2}\Delta_{2})\sigma]\label{witgellmann}
\end{eqnarray}
Hence,
\begin{eqnarray}
&&F(\sigma)\geq Tr(\sigma \vert \delta \rangle \langle \delta \vert)
\label{ineq}
\end{eqnarray}
The r.h.s. may be expressed as 
$\frac{1}{9}Tr((I \otimes I+\frac{3}{2}(\Delta_{2}-\Delta_{1}))\sigma)$ which
using Eq.(\ref{witgellmann}) becomes
$Tr((W^{3}-\frac{1}{3}\Delta_{1})\sigma)$. It follows from Eq.(\ref{ineq}) that
\begin{eqnarray}
Tr[(\frac{1}{3}\Delta_{1}+\frac{1}{3}I-W^{3})\sigma]\geq \frac{1}{3}-F(\sigma)
\end{eqnarray}
Hence, if $\sigma$ is not useful for teleportation , i.e., $F(\sigma)\leq\frac{1}{3}$ \cite{fef1}, then $Tr[(\frac{1}{3}\Delta_{1}+\frac{1}{3}I-W^{3})\sigma]\geq 0$. Thus,
\begin{equation}
W_{3 \otimes 3} = \frac{1}{3}\Delta_{1}+\frac{1}{3}I-W^{3}
\label{telqutrit}
\end{equation}
is indeed a teleportation witness for qutrits.

Now, let us denote by $P_{W_{3 \otimes 3}}$, the set of all product vectors on 
which the expectation value of the witness $W_{3 \otimes 3}$ vanishes, i.e., 
$P_{W_{3 \otimes 3}}=\lbrace \vert e,f \rangle : \langle e,f \vert W_{3 \otimes 3} \vert e,f \rangle =0  \rbrace$. If we consider the product vectors
$K_{1}=\vert 00 \rangle , K_{2}=\vert 11 \rangle , K_{3}= \vert 22 \rangle , K_{4}= (\vert 0 \rangle + \vert 1 \rangle + \vert 2 \rangle)^{\otimes 2},K_{5} = (\vert 0 \rangle + i\vert 1 \rangle) \otimes (\vert 0 \rangle - i\vert 1 \rangle),K_{6}= (\vert 0 \rangle + i\vert 2 \rangle) \otimes (\vert 0 \rangle - i\vert 2 \rangle), K_{7}=(\vert 1 \rangle + i\vert 2 \rangle) \otimes (\vert 1 \rangle - i\vert 2 \rangle), K_{8}=(\vert 0 \rangle - \vert 1 \rangle - \vert 2 \rangle)^{\otimes 2},K_{9}=(\vert 0 \rangle + \vert 1 \rangle - \vert 2 \rangle)^{\otimes 2}$, it is noticed that (i) $\langle K_{i} \vert W_{3 \otimes 3} \vert K_{i}\rangle=0$,
(ii) $K_{i}$'s are linearly independent, $\forall i \in \lbrace 1,2,..9 \rbrace$.
Thus it follows that $P_{W_{3 \otimes 3}}$ spans $C^{3} \otimes C^{3}$. This 
ascertains the optimality of the witness $W_{3 \otimes 3}$ \cite{optwit}.

\subsection{Teleportation witness for qudits}

For general qudit systems the construction of teleportation witnesses from
entanglement witnesses may be undertaken in a manner similar to that shown
above for qubits or qutrits. Utilising the generalized Gell-Mann matrices 
for $d \otimes d$ systems, and retracing the steps of an argument similar to 
that used for qubits and qutrits, one can obtain a teleportation witness for
 qudits as 
\begin{equation}
W_{d \otimes d}=\frac{1}{d}\sum_{j=0}^{d-2}\sum_{k=j+1}^{d-1}(\Lambda_{a}^{jk} \otimes \Lambda_{a}^{jk}) + \frac{1}{d}I-(\vert \Phi \rangle \langle \Phi \vert)^{T_{A}}
\end{equation}
where, $\Lambda_{a}^{jk}= \text{-i}\vert j \rangle \langle k \vert + \text{i} \vert k \rangle \langle j \vert, 0\leq j < k \leq d-1$ and $\vert \Phi \rangle = \frac{1}{\sqrt{d}}\sum_{l=0}^{d-1} \vert ll \rangle$.
Here it may be remarked that there is no general proof of optimality for 
teleportation witness for qudits, but optimality for a given dimension 
needs to be checked in the manner above by considering the set of all
product vectors on which the expectation value of the witness vanishes.

\section{ILLUSTRATIONS AND DECOMPOSITION}

We now consider certain classes of states pertaining to qubits and qutrits, 
which exemplify the action of our constructed witness. Let us first take the 
class of two qubit states with maximally mixed marginals, given by
\begin{equation}
\eta_{mix}=\frac{1}{4}(I \otimes I + \sum_{i=1}^{3}c_{i}\sigma_{i} \otimes \sigma_{i})
\end{equation}   
The expectation value of the witness given by Eq.(\ref{telqubit}) on the above 
state gives
\begin{equation}
Tr(W_{2 \otimes 2}\eta_{mix})=\frac{1}{4}(1+c_{2}-c_{1}-c_{3})
\end{equation}
 implying that for $1+c_{2}-c_{1}-c_{3}<0$, the witness $W_{2 \otimes 2}$ detects the states as useful for teleportation. Since $W_{2 \otimes 2}$ is optimal, this is the largest set of states useful for teleportation in the given class that can be detected by any witness.
Next, we consider the isotropic state in qutrits, given by
\begin{equation}
\eta_{iso}=\alpha \vert \phi_{+}^{3} \rangle \langle \phi_{+}^{3} \vert + \frac{1-\alpha}{9}I
\end{equation}
where, $\vert \phi_{+}^{3} \rangle = \frac{1}{\sqrt{3}}(\vert 00 \rangle + \vert 11 \rangle + \vert 22 \rangle)$ and $-\frac{1}{8}\leq \alpha \leq 1$. Now applying
the witness given by Eq.(\ref{telqutrit}), it is observed that 
\begin{equation} 
Tr(W_{3 \otimes 3}\eta_{iso})=\frac{2-8\alpha}{9}
\end{equation} 
implying that for $\alpha>\frac{1}{4}$, the states are useful for teleportation. Thus, the witness $W_{3 \otimes 3}$ detects all entangled isotropic states as 
useful for teleportation, in conformity with a result already known in the
literature \cite{fef3}. This is a reaffirmation of the optimality of the 
witness $W_{3 \otimes 3}$, as it detects the maximal class of isotropic states 
as useful for teleportation.

The practical use for teleportation witnesses is that they are 
experimentally realizable on account of being hermitian. For qubit systems, 
the decomposition of a proposed teleportation witness in terms of Pauli spin
operators has been shown earlier \cite{ganguly3}. The teleportation witness 
constructed here is expressed in terms of 
generalized Gell-Mann matrices which are hermitian. However, for $d=3$, i.e., 
qutrit systems the teleportation witness can also be expressed in terms of 
spin-$1$ operators \cite{bertlmann} which are the observables $S_{x}, S_{y}, S_{z} ,S_{x}^2 ,S_{y}^2 ,S_{z}^2 ,\{S_{x},S_{y}\},
\{S_{y},S_{z}\}, \{S_{z},S_{x}\} $ of a spin-$1$ system, where
$\overrightarrow{S}=\{S_{x}, S_{y}, S_{z}\}$ is the spin operator
and $\{S_{i},S_{j}\}= S_{i}S_{j}+S_{j}S_{i}$ (with $i,j=x,y,z$)
denotes the corresponding anticommutator. They are given by \cite{bertlmann},
$
S_{x}=\frac{\hbar}{\sqrt{2}}\left(%
\begin{array}{ccc}
  0 & 1 & 0 \\
  1 & 0 & 1 \\
  0 & 1 & 0 \\
\end{array}%
\right),  S_{y}=\frac{\hbar}{\sqrt{2}}\left(%
\begin{array}{ccc}
  0 & \text{-i} & 0 \\
  \text{i} & 0 & \text{-i} \\
  0 & \text{i} & 0 \\
\end{array}%
\right),
S_{z}=\hbar\left(%
\begin{array}{ccc}
  1 & 0 & 0 \\
  0 & 0 & 0 \\
  0 & 0 & -1 \\
\end{array}%
\right)$.
Expressing the witness given by Eq.(\ref{telqutrit}) in terms of spin-$1$ operators, yields
\begin{equation}
W_{3 \otimes 3}= -\frac{2}{9}(I \otimes I)+ \Pi
\end{equation}
where 
\begin{eqnarray}
\Pi &&= \frac{1}{6\hbar^{2}}(S_{y} \otimes S_{y}-S_{z}\otimes S_{z}-S_{x} \otimes S_{x})\nonumber\\
&& + \frac{1}{6\hbar^{4}}(-\lbrace S_{z},S_{x} \rbrace \otimes \lbrace S_{z},S_{x} \rbrace + \lbrace S_{x},S_{y} \rbrace \otimes \lbrace S_{x},S_{y} \rbrace \nonumber\\
&& +  \lbrace S_{y},S_{z} \rbrace \otimes \lbrace S_{y},S_{z} \rbrace) + \frac{2}{3\hbar^{2}}(I \otimes S_{x}^{2} + I \otimes S_{y}^{2}\nonumber\\
&& + S_{x}^{2} \otimes I + S_{y}^{2} \otimes I) - \frac{2}{3\hbar^{4}}(S_{x}^{2} \otimes S_{x}^{2} + S_{y}^{2} \otimes S_{y}^{2})\nonumber\\
&& - \frac{1}{3\hbar^{4}}(S_{x}^{2} \otimes S_{y}^{2} + S_{y}^{2} \otimes S_{x}^{2})
\end{eqnarray}
Thus, for an experimental outcome,
\begin{equation}
\langle W_{3 \otimes 3} \rangle = -\frac{2}{9} \langle I \otimes I \rangle +  \langle \Pi \rangle <0
\end{equation}
one can detect the given unknown state as useful for teleportation.

\section{SUMMARY}

We have presented here a method to construct teleportation witnesses from 
entanglement witnesses for general qudit systems. Optimality of the 
witnesses that we have 
constructed for qubit and qutrit states ensures a broader perspective in the 
sense that a maximal class of entangled states can now be recognized to be 
useful for teleportation. Decomposition of the proposed witness in terms of 
spin operators authenticates its feasibility in experimental detection of
entanglement. The present analysis may be extended in a few
directions. One may seek to 
test the optimality of the witness for two-qudits of any given dimension $d>3$.
Finally, the choice of the entanglement witnesses 
are not limited to the ones we have taken up here, and other entanglement
witnesses may be considered and checked for their viability in the construction
of teleportation witnesses using similar methods.  

{\it Acknowledgements} ASM acknowledges support from the DST Project SR/S2/PU-16/2007.

\end{document}